\documentclass[aps,prl,twocolumn,superscriptaddress]{revtex4-1}
\usepackage{graphicx,color,graphics}
\usepackage{dcolumn} 
\usepackage{amsmath}
\usepackage{amssymb}
\usepackage{amsfonts}
\usepackage{url}
\usepackage{bm}
\usepackage{lipsum}
\usepackage{xcolor}
\makeatletter

\newcommand{\ket}[1]{{{|}{#1}\rangle}}
\newcommand{\bra}[1]{{\langle{#1}{|}}}
\newcommand\blankfootnote[1]{%
  \let\thefootnote\relax\footnotetext{#1}%
  \let\thefootnote\svthefootnote%
}

\newif\ifcmnt
\cmnttrue

\ifcmnt
    \providecommand{\aucmnt}[1]{#1}
\else
    \providecommand{\aucmnt}[1]{}
\fi

\begin{document}

\title{Hyperfine averaging by dynamic decoupling in a multi-ion lutetium clock}
\author{R. Kaewuam}
\affiliation{Centre for Quantum Technologies, National University of Singapore, 3 Science Drive 2, 117543 Singapore}
\author{T. R. Tan}
\affiliation{Centre for Quantum Technologies, National University of Singapore, 3 Science Drive 2, 117543 Singapore}
\affiliation{Department of Physics, National University of Singapore, 2 Science Drive 3, 117551 Singapore}
\author{K. J. Arnold}
\affiliation{Centre for Quantum Technologies, National University of Singapore, 3 Science Drive 2, 117543 Singapore}
\author{S. R. Chanu}
\affiliation{Centre for Quantum Technologies, National University of Singapore, 3 Science Drive 2, 117543 Singapore}
\author{Zhiqiang Zhang}
\affiliation{Centre for Quantum Technologies, National University of Singapore, 3 Science Drive 2, 117543 Singapore}
\author{M. D. Barrett}
\email{phybmd@nus.edu.sg}
\affiliation{Centre for Quantum Technologies, National University of Singapore, 3 Science Drive 2, 117543 Singapore}
\affiliation{Department of Physics, National University of Singapore, 2 Science Drive 3, 117551 Singapore}
\date{\today}
\begin{abstract}
We propose and experimentally demonstrate a scheme which effects hyperfine averaging during a Ramsey interrogation of a clock transition.  The method eliminates the need to average over multiple optical transitions, reduces the sensitivity of the clock to its environment, and reduces inhomogeneous broadening in a multi-ion clock.  The method is compatible with auto-balanced Ramsey spectroscopy, which facilitates elimination of residual shifts due to imperfect implementation and ac stark shifts from the optical probe.  We demonstrate the scheme using correlation spectroscopy of the $^1S_0\leftrightarrow {}^3D_1$ clock transition in a three-ion Lu$^+$ clock.  From the demonstration we are able to provide a measurement of the $^3D_1$ quadrupole moment, $\Theta({}^3D_1)=0.634(9) e a_0^2$.
\end{abstract}
\maketitle

Optical atomic clocks have seen rapid advances over the last decade with a number of systems now reaching fractional uncertainties near to $\sim 10^{-18}$ \cite{Ludlow2015,Nicholson2015,mcgrew2018atomic,Chou2010,Huntemann2016}.  Singly ionized lutetium ($^{176}$Lu$^+$) is a relatively new clock candidate with a number of attractive features for clock applications \cite{barrett2015developing,Arnold2018,tan2019suppressing}.  It provides three independent clock transitions allowing consistency checks on error budgets through frequency comparisons within the one system.  Two of the transitions allow long interrogation times relevant to current state-of-the-art lasers, and have atomic properties relevant to clock performance that compare favourably to other leading clock candidates \cite{Arnold2018}.  As with other ion-based clocks, single-ion operation limits fractional instability to $\sim 10^{-15}/\sqrt{\tau}$, for which averaging times ($\tau$) of several days or even weeks are required to reach $10^{-18}$ resolution \cite{sanner2019optical}.  For this reason, multi-ion clock implementations are of interest \cite{Pyka2014,Arnold2015,shaniv2019quadrupole,tan2019suppressing}.  Recent work illustrated the feasibility of a $^{176}$Lu$^+$ multi-ion approach, by demonstrating suppression of inhomogeneous broadening to the level of $10^{-17}$ and the potential for clock interrogation times $\gtrsim 10\,\mathrm{s}$.  However, averaging over hyperfine states is still required to make full use of the advantages $^{176}$Lu$^+$ has to offer \cite{barrett2015developing}.

All optical clock implementations use some form of averaging to cancel various systematic shifts, be it averaging over Zeeman pairs \cite{dube2013evaluation}, hyperfine states \cite{barrett2015developing}, or orthogonal orientations of the magnetic field \cite{Itano2000}.  In a multi-ion crystal, individual systematic shifts become a source of inhomogeneous broadening, which can diminish the efficacy of averaging particularly for long interrogation times; as the interrogation time increases, individual shifts become increasingly resolved, which distorts the line shape of a multi-ion spectroscopy signal and shifts the line center away from the mean.   Hence, broadening mechanisms must be well characterized and controlled, which can be a significant challenge in an ensemble of ions.  Nevertheless progress towards multi-ion operation has been made: precision engineering of the ion trap has allowed the control of excess-micromotion (EMM) shifts to the $10^{-19}$ level over millimeter length scales \cite{Keller2019}; precise alignment of the magnetic field has demonstrated suppression of tensor shifts \cite{tan2019suppressing}; and dynamic decoupling during interrogation has also demonstrated suppression of inhomogeneous broadening \cite{shaniv2019quadrupole}.

Dynamic decoupling is an attractive approach to suppressing inhomogeneous broadening, as it also reduces the sensitivity of the clock frequency to electromagnetic fields, and simplifies the averaging by eliminating the need to probe multiple transitions.  However, the recent demonstration in $^{88}$Sr$^+$ \cite{shaniv2019quadrupole} is not directly applicable to $^{176}$Lu$^+$.  Instead we use a simple scheme in which population is transferred between hyperfine states during a Ramsey interrogation so that each upper state is sampled for an equal duration of time.  This effectively implements hyperfine averaging within a single interrogation sequence and can be interpreted as a form of dynamic decoupling.  We start with a theoretical description of the technique and then give an experimental demonstration using optical correlation spectroscopy.

\begin{figure*}
\includegraphics[width=\linewidth]{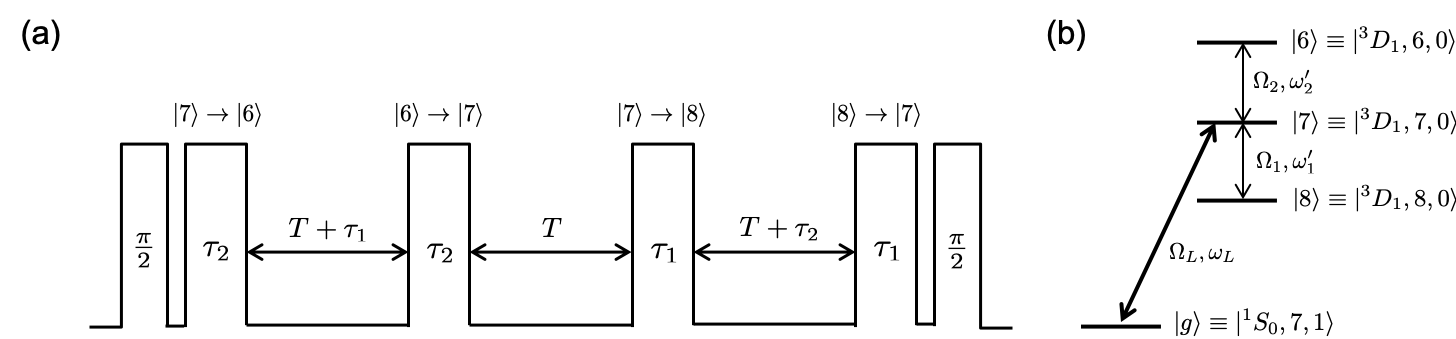}
\caption{Dynamic decoupling used to effect hyperfine averaging.  (a) Pulse sequence used. The $\pi/2$-pulses provide Ramsey spectroscopy of the $\ket{g}\leftrightarrow\ket{7}$ transition.  Pulses labelled $\tau_1$ and $\tau_2$ are microwave $\pi$-pulses driving $\ket{8}\leftrightarrow\ket{7}$ and $\ket{6}\leftrightarrow\ket{7}$ respectively. (b) Level scheme for the states involved.}
\label{Ramsey}
\end{figure*}

Consider a Ramsey interrogation of the $^1S_0$ to $^3D_1$ transition using the sequence shown in Fig.~\ref{Ramsey}, in which microwave transitions within the interrogation time are used to effect hyperfine averaging.  The ground state and upper $m=0$ states are denoted $\ket{g}$ and $\ket{F}$, respectively.  A laser of frequency $\omega_L$ drives the optical transition $\ket{g}\leftrightarrow\ket{7}$ with a coupling strength $\Omega_L$.  Microwave frequencies and coupling strengths are denoted $\omega'_k$ and $\Omega_k$, respectively, with $k=1$ corresponding to the transition $\ket{8}\leftrightarrow\ket{7}$ and $k=2$ to $\ket{7}\leftrightarrow\ket{6}$.  Although Fig.~\ref{Ramsey} shows a sequence relevant to the optical addressing of $\ket{g}\leftrightarrow\ket{7}$, any upper state, $\ket{F}$, may be used with an appropriate reordering of the microwave pulses.  In the general case, the Hamiltonian in the interaction picture is
\begin{multline}
H_I=\frac{\Omega_L}{2}\left(e^{-i(\omega_L-\omega_F)t}\ket{F}\bra{g}+e^{i(\omega_L-\omega_F)t}\ket{g}\bra{F}\right)\\
+\frac{\Omega_2}{2}\left(e^{-i\Delta_2 t}\ket{6}\bra{7}+e^{i\Delta_2 t}\ket{7}\bra{6}\right)\\
+\frac{\Omega_1}{2}\left(e^{-i\Delta_1 t}\ket{7}\bra{8}+e^{i\Delta_1t}\ket{8}\bra{7}\right),
\end{multline}
where we have kept only near resonant terms.  Upper state energies $\hbar\omega_F$ are taken relative to the zero-point energy of the ground-state, and $\Delta_k=\omega'_k-\omega_k$ are the microwave detunings relative to their respective resonances $\omega_1=\omega_7-\omega_8$ and $\omega_2=\omega_6-\omega_7$.

In what follows, it is convenient to introduce the signed frequency combinations
\[
\bar{\omega}_6=\frac{2\omega_2+\omega_1}{3},\quad\bar{\omega}_7=-\frac{\omega_2-\omega_1}{3},\quad \bar{\omega}_8=-\frac{2\omega_1+\omega_2}{3},
\]
and counterparts, $\bar{\omega}'_F$, defined in a similar way with $\omega'_k$ replacing $\omega_k$.  The frequencies $\bar{\omega}_F$ are the offsets of a given level from the hyperfine averaged frequency, $\omega_0$. That is,
\begin{equation}
\omega_0=\frac{\omega_6+\omega_7+\omega_8}{3}=\omega_F-\bar{\omega}_F.
\end{equation}
With these definitions at hand, we transform to a rotating frame using $U=e^{i H_0 t}$ where 
\begin{equation}
H_0=-\delta \ket{g}\bra{g}+\sum_F\Delta_F\ket{F}\bra{F},
\end{equation}
with the parameters $\delta$ and $\Delta_F$ conveniently set to
\begin{equation}
\Delta_F=\bar{\omega}'_F-\bar{\omega}_F=\omega_0+\bar{\omega}'_F-\omega_F,
\end{equation}
and
\begin{equation}
\delta=\omega_L-\omega_F-\Delta_F=\omega_L-\bar{\omega}_F'-\omega_0.
\end{equation}
In the rotating frame we then have
\begin{multline}
H=\delta \ket{g}\bra{g}-\sum_F\Delta_F\ket{F}\bra{F}+\frac{\Omega_L}{2}\left(\ket{F}\bra{g}+\ket{g}\bra{F}\right)\\
+\frac{\Omega_2}{2}\left(\ket{6}\bra{7}+\ket{7}\bra{6}\right)+\frac{\Omega_1}{2}\left(\ket{7}\bra{8}+\ket{8}\bra{7}\right).
\end{multline}
We can now consider the phase accumulated in a Ramsey sequence including two microwave $\pi$-pulses from both sources as required to cycle through all states.  We denote the microwave $\pi$-pulse times by $\tau_k=\pi/(2\Omega_k)$ and consider dwell times of $T+\tau_1$, $T,$ and $T+\tau_2$ for $\ket{6}$, $\ket{7},$ and $\ket{8}$ respectively.    Starting from the initial state $\ket{\psi_0}=\ket{g}-i\ket{F}$ arising from the first optical Ramsey pulse, the ground state accumulates a phase $-3\delta (T+\tau_1+\tau_2)$ and the upper state accumulates
\begin{multline}
T\sum_F \Delta_F+\tau_1\Delta_6+\tau_2\Delta_8\\
+(\Delta_8+\Delta_7)\tau_1+(\Delta_7+\Delta_6)\tau_2,
\end{multline}
where terms on the first line originate from the dwell time in each upper state, and the remaining terms originate from the $\pi$-pulses. Since $\sum_F \Delta_F=0$, the phase shift on the upper state is zero, and the central Ramsey fringe satisfies $\delta=0$. From the definition of $\delta$, this is the shift relative to $\omega_0+\bar{\omega}'_F$.  Down-shifting the laser by the well defined microwave offset $\bar{\omega}'_F$, then gives the hyperfine averaged frequency, $\omega_0$, free of any tensor shifts and associated inhomogeneous broadening. 

To understand the phase evolution during the microwave pulse consider the two level system with Hamiltonian
\begin{equation*}
\begin{pmatrix} \delta_1 & \Omega \\ \Omega & \delta_2 \end{pmatrix}=\begin{pmatrix} \frac{\delta_1+\delta_2}{2} & 0 \\ 0 & \frac{\delta_1+\delta_2}{2} \end{pmatrix}+\begin{pmatrix} \frac{\delta_1-\delta_2}{2} & \Omega\\ \Omega & -\frac{\delta_1-\delta_2}{2} \end{pmatrix}
\end{equation*}
The first term clearly commutes with the second so the unitary evolution factors.  The second term acquires no phase, whereas the first term gains a global phase.  In a conventional Ramsey interrogation a global phase has no influence on the spectroscopy.  However, using microwave pulses within the Ramsey time, only applies that phase to the upper states. 

The additional dwell time in $\ket{6}$ and $\ket{8}$ can be intuitively understood.  During a microwave $\pi$-pulse between two upper states, each state is occupied half the time.  Thus, during a $\pi$-pulse, we obtain a phase shift corresponding to the average shift from two states: $\ket{6}$ and  $\ket{7}$ for one pulse, and $\ket{7}$ and  $\ket{8}$ for the other.  For all pulses used, $\ket{7}$ is over-represented or, equivalently, $\ket{6}$ and $\ket{8}$ are under-represented.  This is then easily compensated by increasing the dwell time in $\ket{6}$ and $\ket{8}$ by the appropriate $\pi$-pulse time.

To demonstrate the method, we use a combination of microwave and optical spectroscopy in a three-ion optical clock as described in our recent work \cite{tan2019suppressing}.  We first align the magnetic field along the trap axis, which maximises inhomogeneous broadening between the ions. We then show the broadening is eliminated by the decoupling protocol.  

The magnetic field is aligned with respect to a $\pi$-polarized optical pumping beam, which addresses the $^3D_1, F=7$ to $^3P_0, F'=7$ transition at $646\,\mathrm{nm}$.  Maximising depumping times out of $\ket{7}$ by the optical pumping beam ensures the magnetic field is well aligned to the laser polarisation,  which geometry constrains to about $\pm 5^\circ$ with respect to the crystal axis.  However the actual angle is not crucial for our demonstration.

\begin{figure*}
\includegraphics[width=\linewidth]{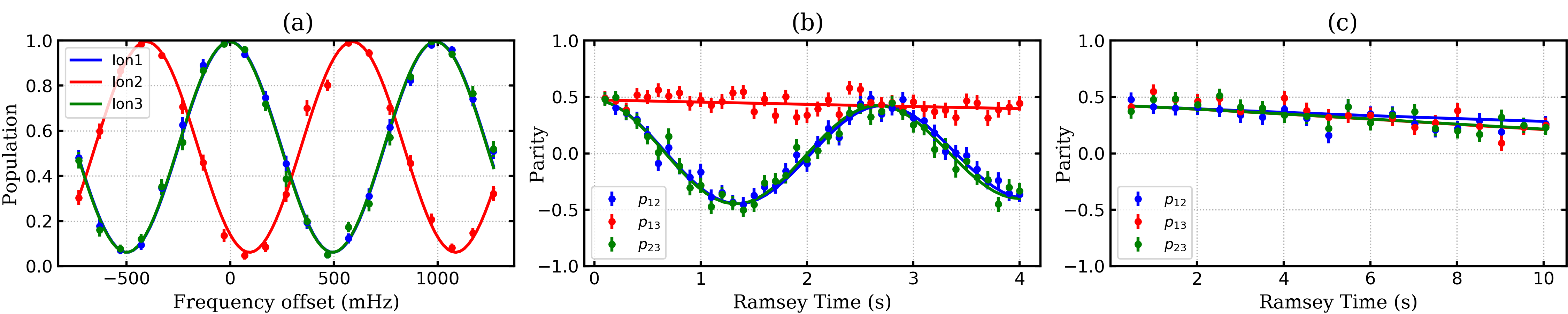}
\caption{(a) Microwave Ramsey spectroscopy of the $\ket{{}^3D_1,8,0}\leftrightarrow\ket{{}^3D_1,7,0}$ transition using a $1\,\mathrm{s}$ Ramsey time.  Frequencies on the horizontal axes are relative to the measured hyperfine splitting of $\sim10.49\,\mathrm{GHz}$ of the outer ions (ions 1\&3).  The center fringe for the middle ion (ion 2) is offset by $\approx 0.6\,\mathrm{Hz}$. (b) Correlation spectroscopy on the $\ket{{}^1S_0,7,1}\leftrightarrow\ket{{}^3D_1,7,0}$ transition without dynamic decoupling,  (c) Correlation spectroscopy on the $\ket{{}^1S_0,7,1}\leftrightarrow\ket{{}^3D_1,7,0}$ transition using the decoupling sequence shown in Fig.\ref{Ramsey}.} 
\label{Spectroscopy}
\end{figure*}

Ramsey spectroscopy of the $\ket{8}\leftrightarrow\ket{7}$ transition is carried out to characterise broadening arising from differential tensor shifts between the ions.  As in previous work \cite{tan2019suppressing}, the most significant contribution is due to the coupling of the quadrupole moment to the Coulomb field from neighbouring ions, with  axial micromotion arising from the end-cap design of the trap contributing at the $1\%$ level from the associated ac-Stark shift.  Provided micromotion is well compensated for the middle ion, both effects shift the outer ions equally with respect to the middle ion as is evident from the data shown in Fig.~\ref{Spectroscopy}(a).      

To demonstrate the averaging protocol, we use correlation spectroscopy of the $\ket{g}\equiv\ket{^1S_0,7,1}\leftrightarrow\ket{7}$ transition both with and without decoupling.  As discussed elsewhere \cite{tan2019suppressing,Chwalla2007,Chou2011}, correlation spectroscopy enables frequency comparison between pairs of ions beyond the radiation's coherence time.  Under the assumption of a uniform distribution of phases,  the expectation value of the parity operator $p_{ij}=\langle\sigma_{z,i}\sigma_{z,j}\rangle$ is
\begin{equation}
p_{ij} = \frac{p_c}{2} \cos\left(2\pi \Delta f_{ij} T\right),
\label{ParityInCoherentEq}
\end{equation}
where $p_c$ characterizes the loss of relative coherence between the two atomic oscillators under investigation, and $\Delta f_{ij}$ is the difference in their resonant transition frequencies.  

Results of correlation spectroscopy without decoupling is shown in Fig.~\ref{Spectroscopy}(b).  As expected, there is an oscillation in the correlation signals $p_{12}$ and $p_{23}$ and no significant oscillation in $p_{13}$.  The differential tensor shift between ions inferred from the correlation signals is $589.0(3.2)\,\mathrm{mHz}$ which is consistent with $592.2(4.5)\,\mathrm{mHz}$ measured from the microwave Ramsey spectroscopy, Fig.~\ref{Ramsey}(a).  The small frequency difference in the correlation signals $p_{12}$ and $p_{23}$ is due to a $113\,\mathrm{\mu T/m}$ magnetic field gradient and the $\sim3.4\mathrm{kHz/mT}$ Zeeman sensitivity of $\ket{g}$.  The gradient was independently measured via microwave correlation spectroscopy of the $\ket{^3D_1,6,-1}\leftrightarrow\ket{7}$ transition as demonstrated in \cite{tan2019suppressing}.  With the decoupling sequence shown in Fig.~\ref{Ramsey}, oscillations in $p_{12}$ and $p_{23}$ are no longer present as shown in Fig.~\ref{Spectroscopy}(c).  As seen in our previous work \cite{tan2019suppressing}, there is a noticable decay that cannot be explained by the expected frequency differences from micromotion and magnetic field gradients.  To allow for possible dephasing, we fit the data to Eq.~\ref{ParityInCoherentEq} using $p_c=p_0^2\exp(-t/T_c)$, with $p_0$ and $T_c$ as free parameters, and $\Delta f_{ij}$ set to the measured values.   The resulting dephasing time of $T_c=24(3)\,\mathrm{s}$ is consistent with the $T_c=27(6)\,\mathrm{s}$ found in our previous work \cite{tan2019suppressing}.

Due consideration should be given to the microwave $\pi$-pulse times used.  If the pulses are too long, they could resolve the shifts of individual ions. This is unlikely to be a problem for practical clock operation, as the magnetic field would be first aligned to suppress inhomogeneous broadening as demonstrated in \cite{tan2019suppressing}.  With shorter pulse times, one must account for ac Zeeman shifts arising from off-resonant microwave couplings, particularly from the $\sigma^\pm$ components of the microwave fields.  If the $\sigma^\pm$ components are balanced, shifts will largely cancel.  More generally, ac Zeeman shifts will result in a clock shift of
\begin{equation}
\label{shift}
\delta=\frac{(\Delta_{1,7}+\Delta_{1,8})\tau_1+(\Delta_{2,6}+\Delta_{2,7})\tau_2}{3(T+\tau_1+\tau_2)},
\end{equation}
where $\Delta_{k,F}$ is the ac Zeeman shift of $\ket{F}$ when $\Omega_k$ is applied.  This shift has a similar form as that from a probe-induced ac stark shift when using Ramsey spectroscopy.  Consequently, auto-balanced Ramsey spectroscopy would simultaneously eliminate both optical probe- and microwave-induced shifts \cite{sanner2018autobalanced,yudin2018generalized}.  For this experiment, we measured the $\pi$ times for all $\Delta m=0,\pm 1$ transitions from $\ket{7}$ and the results are tabulated in table~\ref{uwaveShift}.  Scaling the results to the $10\,\mathrm{ms}$ $\pi$-times used in this experiment, we estimate the shift from Eq.~\ref{shift} to be $\sim 2\times 10^{-20}$ relative to the clock frequency.    
\begin{table}
 \caption{Measured $\pi$-times for transitions $\ket{^3D_1,7,0}\leftrightarrow\ket{^3D_1,F,m}$ for $F=6,8$ and $m=0,\pm1$; $\pi$-times for transitions $\ket{^3D_1,7,\pm1}\leftrightarrow\ket{^3D_1,F,0}$ can be determined by scaling with appropriate Clebsch-Gordan coefficients.}
 \label{uwaveShift}
\begin{ruledtabular}
\begin{tabular}{c c c c}
 \hspace{0.5cm}Transition &  \hspace{0.5cm} $m=-1$ & \hspace{0.25cm} $m=0$ & \hspace{0.25cm} $m=1$ \hspace{0.5cm} \\
 \hline
 \vspace{-0.3cm}\\
\hspace{0.5cm} $F=6$ & \hspace{0.25cm} 7\,ms & \hspace{0.25cm} 3.5\,ms & \hspace{0.25cm} 9\,ms \hspace{0.5cm} \\
\hspace{0.5cm} $F=8$ & \hspace{0.25cm} 23\,ms & \hspace{0.25cm} 3.5\,ms & \hspace{0.25cm} 20\,ms \hspace{0.5cm}
 \end{tabular}
 \end{ruledtabular}
 \end{table}

Our demonstration has been implemented on the $\ket{^1S_0,7,1}\leftrightarrow\ket{^3D_1,7,0}$ transition due to practical considerations of our current laser set up.  However, the technique can be applied starting from any upper $F$ level, with an appropriate change to the microwave offset $\bar{\omega}'_F$.  Starting from the upper $F=8$ level, for example, would allow the $\ket{^1S_0,7,0}\leftrightarrow\ket{^3D_1,8,0}$ transition to be used eliminating the linear Zeeman sensitivity of the ground state.  This leaves only the quadratic Zeeman shifts of the upper $m=0$ states with estimated sensitivities of $2.27$, $-0.12$, and $-2.15\,\mathrm{mHz/\mu T^2}$ for $\ket{6}$, $\ket{7}$, and $\ket{8}$ respectively, and just $-4.7\,\mathrm{\mu Hz/\mu T^2}$ for the hyperfine-averaged clock frequency \cite{paez2016atomic}.  Such sensitivities require  rather modest experimental control of spatial and temporal variations of the ambient magnetic field, even for the long interrogation times made possible with current laser technology.

Note that the differential shift between the outer and inner ions is almost completely determined by the quadrupole shift induced by the neighbouring ions: axial micromotion contributes at the few mHz level and all static trapping fields appear common mode to the ions.  Consequently, the observed frequency differences in Fig.~\ref{Spectroscopy}(a) and (b) can provide a reasonable estimate of the quadrupole moment.  From \cite[Eq.~1]{tan2019suppressing}, the quadrupole contribution to the frequency difference in Fig.~\ref{Spectroscopy}(a) is given by,
\begin{equation}
h\Delta f = \frac{14}{25}\left(3 \cos^2 \theta - 1\right) \Theta(^3D_1) \frac{m \omega_z^2}{e},
\end{equation}
where $e$ is the elementary charge, $m$ is the mass of the ion, $\theta$ is the angle between the applied dc magnetic field and trap axis, and $\omega_z$ is the axial trap frequency.  In this expression, the value of $\Theta(^3D_1)$ is specified in accordance with the definition given in \cite{Itano2000}.  Using the measured trap frequency of $\omega_z=2\pi\times 131.66(1)\,\mathrm{kHz}$, a micromotion contribution of $4.7(2)\,\mathrm{mHz}$ inferred from Raman sideband spectroscopy, and allowing $|\theta|<5^\circ$, gives an estimated quadrupole moment of $\Theta(^3D_1) = 0.634(9)\,ea^2_0$ where $a_0$ is the Bohr radius.  This value is comparable to the theoretical estimate of $\Theta(^3D_1) = 0.655\,ea^2_0$ given in \cite{Safronova2018}, taking into account their choice of convention and omitted sign.

In summary, we have proposed and demonstrated a simple technique to effect hyperfine averaging during Ramsey interrogation of a $^{176}$Lu$^+$ clock transition.  The technique can be viewed as a form of dynamic decoupling, which eliminates dominant sources of inhomogeneous broadening in a multi-ion clock implementation.  The level of suppression demonstrated is limited by the maximum differential quadrupole shift and interrogation time currently achievable.  Even with this limitation, we expect tensor shifts to be suppressed well below $10^{-18}$ when using the technique in conjunction an optimal field alignment as demonstrated in \cite{tan2019suppressing}.

The implementation can be readily adapted so as to utilize only $m=0$ states providing improved resilience against magnetic field noise.  At our current operating field of $\sim0.1\,\mathrm{mT}$, component $m=0$ states involved in the interrogation are two orders of magnitude less sensitive to magnetic field noise compared to the stretch states used in Al$^+$, for example.  This is an important practical consideration for achieving long interrogation times.  In addition, the technique is compatible with auto-balanced Ramsey spectroscopy, so that probe induced ac Stark shifts can also be eliminated \cite{sanner2018autobalanced,yudin2018generalized}.

We have also provided a straightforward means to measure the quadrupole moment for the $^3D_1$ level utilizing the differential quadrupole shift induced by neighbouring ions.  All applied static electric fields including stray fields appear common mode so that characterization of these fields is unnecessary.  Our current estimate is limited by the ability to characterize the angle of the magnetic field with respect to the ion crystal and a small amount of axial micromotion.  Alignment of the field could be improved by maximizing the tensor shift with respect to bias coil currents and micromotion diminished by increasing the axial confinement and lowering the rf drive voltage.  We anticipate that an order of magnitude improvement in the measurement precision should be possible with these improvements.

\begin{acknowledgements}
This work is supported by the National Research Foundation, Prime Ministers Office, Singapore and the Ministry of Education, Singapore under the Research Centres of Excellence programme. This work is also supported by A*STAR SERC 2015 Public Sector Research Funding (PSF) Grant (SERC Project No: 1521200080). T. R. Tan acknowledges support from the Lee Kuan Yew post-doctoral fellowship.
\end{acknowledgements}
\bibliography{Decoupling}
\bibliographystyle{unsrt}
\end{document}